# Cloud computing security using encryption technique

Geethu Thomas geethu.thomask@gmail.com

Prem Jose V premjosev@gmail.com

P.Afsar mhdafsar@gmail.com

***Abstract*— Cloud Computing has been envisioned as the next generation architecture of IT Enterprise. The Cloud computing concept offers dynamically scalable resources provisioned as a service over the Internet. Economic benefits are the main driver for the Cloud, since it promises the reduction of capital expenditure and operational expenditure. In order for this to become reality, however, there are still some challenges to be solved. Most important among these are security and trust issues, since the user's data has to be released to the Cloud and thus leaves the protection sphere of the data owner. In contrast to traditional solutions, where the IT services are under proper physical, logical and personnel controls, Cloud Computing moves the application software and databases to the large data centers, where the management of the data and services may not be fully trustworthy. This unique attribute, however, poses many new security challenges which have not been well understood. Security is to save data from danger and vulnerability. There are so many dangers and vulnerabilities to be handled. Various security issues and some of their solution are explained and are concentrating mainly on public cloud security issues and their solutions. Data should always be encrypted when stored (using separate symmetric encryption keys) and transmitted. If this is implemented appropriately, even if another tenant can access the data, all that will appear is gibberish. So a method is proposed such that we are encrypting the whole data along with the cryptographic key.**

## I. INTRODUCTION

Transformation of computing to services which are customerised and delivered like traditional utilities (water, gas and electricity) depends on computing paradigms such as cluster computing grid computing and recently cloud computing [1, 2] Cloud computing as a utility can transform and attract IT (information technology) services. This innovative idea reduces capital outlays as well as operation costs. Due to this potential capacity cloud computing is a fastest developing field in IT sector. Cloud computing is defined as the delivery of application as services over internet using the software and hardware facility of the service providers which can be either called as Software as a Service (SaaS), Infrastructure as a service (IaaS) or Platform as a Service (PAS) [1]. The hardware and software part forms the *cloud* which is generally called as *public cloud* were services offered in a pay as you use manner; comes under *utility computing*. On the other hand *private cloud* refers the restricted access to general public while full access to that organization/business that avail such facility. Cloud computing is the sum total of SaaS and *utility computing* facility were data centers (small and medium) are excluded while people can be either users or providers of the former aforesaid facility. Transformation of computing world towards development of software as services for a vast group rather targeting individual computers [2].

Business applications capacities are offered by cloud computing services accessed over a network were customers are being charged by the service providers for the services being availed. Cloud computing technology delivers all the IT functionalities and dramatically reduces the upfront costs of computing which may give the cutting edge to the companies [3]. As a part of Total Quality Management (TQM), redundancy and reliability; providers especially Amazon, Google, Salesforce, International Business Management (IBM) and Microsoft have launched data centers for cloud computing around the globe [2]. The foremost milestone towards the goal of achievement of *utility computing* considered to be the vision of $21^{st}$ century was the implementation of Advanced Research Project Agency Network (ARPANET) which later spread its popularity as the World Wide Web (WWW) and internet [4]. Cloud computing combines the convergence of IT efficiency as well as business agility with a real time response to the user requirements [3].

In counterpart to cloud computing, the other widely explored computing paradigms include cluster computing and grid computing. Grid computing enables resource sharing between geographically distributed resources with inspirations from electrical power grids prospective. Cluster computing comprises parallel and inter-connected networks with group of computers working as single integrated computing resources. Google search trends for cloud computing, grid computing and cluster computing from July 2010 till January 2013 is shown in figure 1.

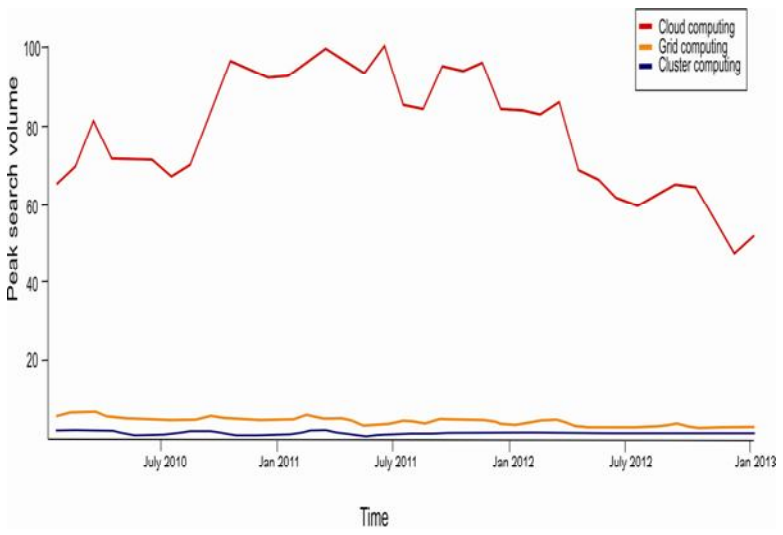

Figure 1. Google search trends for cloud computing, grid computing and cluster computing from July 2010 till January 2013.

With the evolution of cloud computing three core technologies such as virtualisation, multitenancy and web services are rapidly emerging. Virtualisation hides the physical characteristics of a computing platform where multitenancy, allows instances of application software for multiple clients. A web service provides a software system designed to support interoperable machine to machine interaction over a network [3]. The stakeholders in cloud computing is totally different from the traditional computing and involves consumers, providers, enablers and regulators.

In order to reduce capital expenditure and operational expenditure, there are some challenges to be addressed. Amongst these are security and trust issues, since the users data has to be released to the Cloud and thus leaves the protection sphere of the data owner. According to International Data Corporation (IDC) Security is ranked first as the greatest challenge or issue of cloud computing. Experience shows that attacks may never be completely prevented or detected accurately and on time.

The Cloud Security Alliance is a non-profit organization formed to promote the use of best practices for providing security assurance within cloud computing. As more and more information on individuals and companies is placed in the cloud, concerns are beginning to grow about just how safe an environment it is. This paper discusses security issues, requirements and challenges that cloud service providers (CSP) face during cloud engineering and some solutions to mitigate them. It needs some form of standardization (e.g. Information Technology Infrastructure Library -ITIL, Open Virtualization Format (OVF)) so that the market can evolve and thrive. Standards should allow clouds to interoperate and communicate with each other no matter which vendor provides cloud services. It also highly recommends OVF standard as vendor and platform independent, open, secure, portable, efficient and extensible format for the packaging and distribution of software to be run in virtual machines. For the protection of data privacy, sensitive data has to be encrypted before outsourcing, which makes effective data utilization a very challenging task. For maintaining security in the cloud we have to consider some of the issues addressed and its solution. There are some standards and agreements that should allow clouds to interoperate and communicate with each other no matter which vendor provides cloud services. An encryption method is introduced so that we can store our data securely.

Concern among big cooperate companies about handling their operations through another firm and bankruptcy of cloud providers especially in a shrinking economy. Security also a serious concern among IT executives followed by performance and reliability [16]. Lack of standards especially International Organizations for Standards are still missing in cloud services which may reduce its acceptability. The launch of EuroCloud is a typical example were standards are implemented and being checked for the safeguarding the interest of clients throughout European Union (EU) [3].

## II. EXISTING WORK

Cloud computing is a natural evolution of the widespread adoption of virtualization, service-oriented architecture and utility computing. Details are abstracted from consumers, who no longer have need for expertise in, or control over, the technology infrastructure ”in the cloud” that supports them.

The relative security of cloud computing services is a contentious issue which may be delaying its adoption. Issues barring the adoption of cloud computing are due in large part to the private and public sectors unease surrounding the external management of security based services. Organizations have been formed in order to provide standards for a better future in cloud computing services. One organization in particular, the Cloud Security Alliance is a non-profit organization formed to promote the use of best practices for providing security assurance within cloud computing. We categorize the security concerns as: traditional security, availability, third-party data control [7].

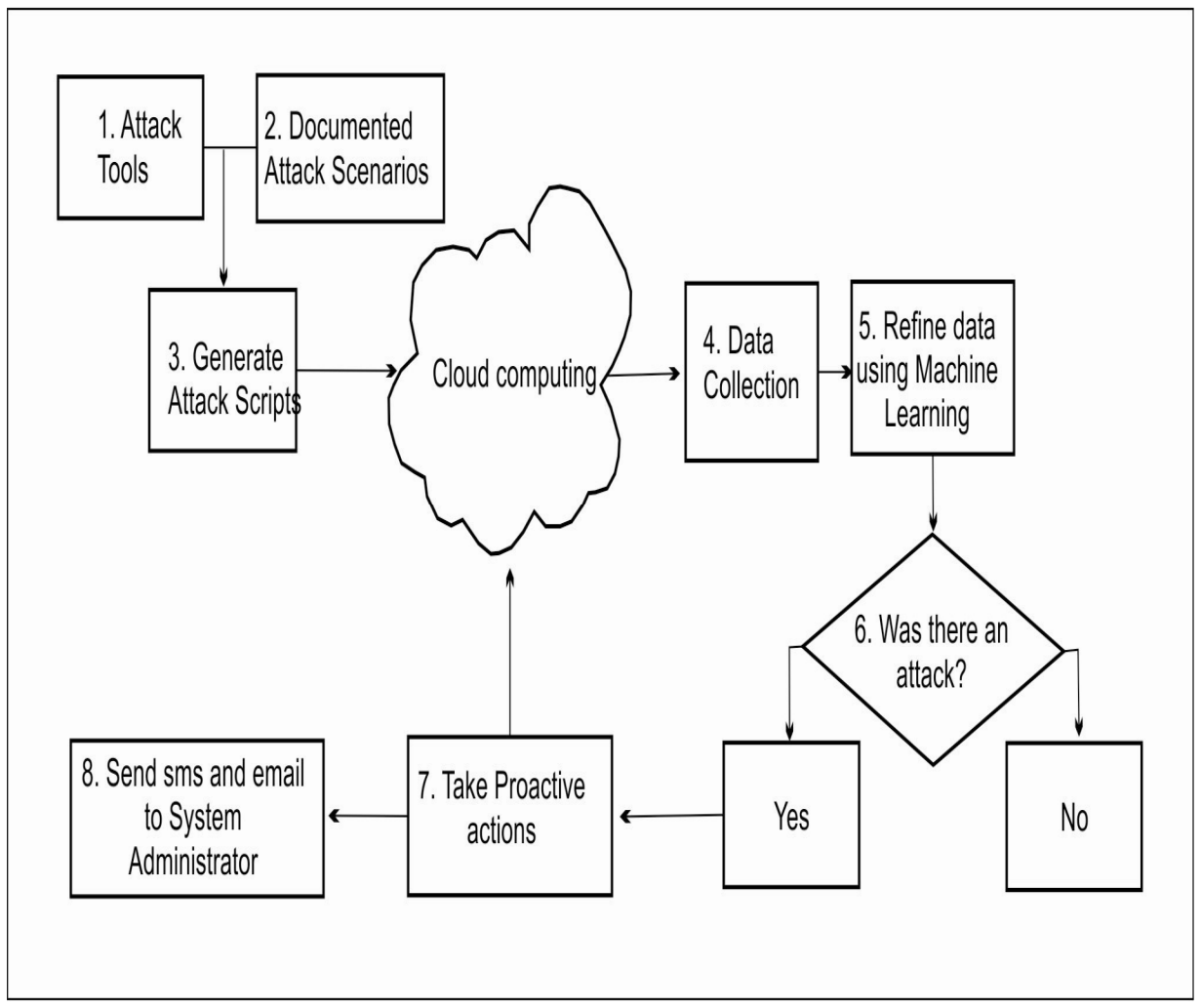

**Figure 12.** Attack detection and proactive resolution in a single cloud environment. Adapted from [6].

1. Traditional Security:

These concerns involve computer and network intrusions or attacks that will be made possible or at least easier by moving to the cloud. Cloud providers respond to these concerns by arguing that their security measures and processes are more mature and tested than those of the average company. Cloud provider vulnerabilities could be platform-level, such as an SQL-injection or cross-site scripting vulnerability in salesforce.com, phishes and other social engineers have a new attack vector. The cloud user must protect the infrastructure used to connect and interact with the cloud, a task complicated by the cloud being outside the firewall in many cases. The enterprise authentication and authorization framework does not naturally extend into the cloud. Potential vulnerabilities in the hypervisor or VM technology used by cloud vendors are a potential problem in multi-tenant architectures. Investigating inappropriate or illegal activity may be difficult in cloud computing because logging and data for multiple customers may be co-located may also be geographically spread across an ever-changing set of hosts and data centers. Solution is to get a contractual commitment to support specific forms of investigation

2. Availability:

These concerns center on critical applications and data being available. Well-publicized incidents of cloud outages include Gmail (one-day outage in mid-October 2008), Amazon S3 (over seven-hour downtime on July 20, 2008). As with the Traditional Security concerns, cloud providers argue that their server uptime compares well with the availability of the cloud user’s own data centers. Cloud services are thought of as providing more availability, but perhaps not - there are more single points of failure and attack. Assurance of computational integrity is another problem.

3. Third-party data control:

The legal implications of data and applications being held by a third party are complex and not well understood. There is also a potential lack of control and transparency when a third party holds the data. Part of the hype of cloud computing is that the cloud can be implementation independent, but in reality regulatory compliance requires transparency into the cloud. Audit difficulty is another side effect of the lack of control in the cloud. Is there sufficient transparency in the operations of the cloud provider for auditing purposes? Due diligence , that is if served a subpoena or other legal action, can a cloud user compel the cloud provider to respond in the required timeframe, contractual obligations, data Lock-in and transitive nature are another possible concerns. Trusted third party can be relied for (1) low and high level of confidentiality (2) service and client authentication (3) creation of security domains (4) cryptographic separation of data (5) certificate based authorization.

4. Low and high level of confidentiality:

Threat of data modification and data interruption is a serious issue in cloud network. Public Key Infrastructure (PKI) enables IPSec of SSL for secure connections. IPSec provide confidentiality and authenticity while SSL protocol generate end to end encryption and an encrypted communication channel between client and server. Communications are protected between user and host but also from host to host. IPSec is compatible with any application and requires IPSec client while SSL is built into every browser.

5. Server and client authentication:

Certifying agencies are required for certifying physical infrastructure servers, virtual servers, environment users and network devices. A certification authority builds the necessary strong credentials for all physical and virtual entities in the cloud.

6. Creation of security domains:

Federation clouds are associated groups of legal entities that share agreed policies and legal frame work across different organizations.

7. Intrusion detection and prevention system:

The integrity and availability of systems need to safeguard against a number of threats which include hackers, rival competitors, terrorists and foreign governments. The growth of wired as well wireless communication networks force the clouds to be secured through firewalls, intrusion detection and prevention system, encryption and authentication . Intrusion detection and prevention systems (IDPS) can early detect the malicious activity and prevent the serious damage to the cloud. IDPS can be used as forensic evidence which can used in legal proceedings. False alarm generation is also associated with IDPS which often disrupt information availability. The wide distributed network and open structure of the cloud make it a good choice for intruders.

## III. ISSUES

The major gaps in cloud computing are availability of services, data lock-in, data confidentiality and auditability, data transfer bottlenecks, performance unpredictability, scalable storage, bugs and software licensing. Cloud computing is associated with the tradeoffs between cost and security. The cloud computing security aspects can be broadly classified into three catergories; security categories, security in service and security dimensions.

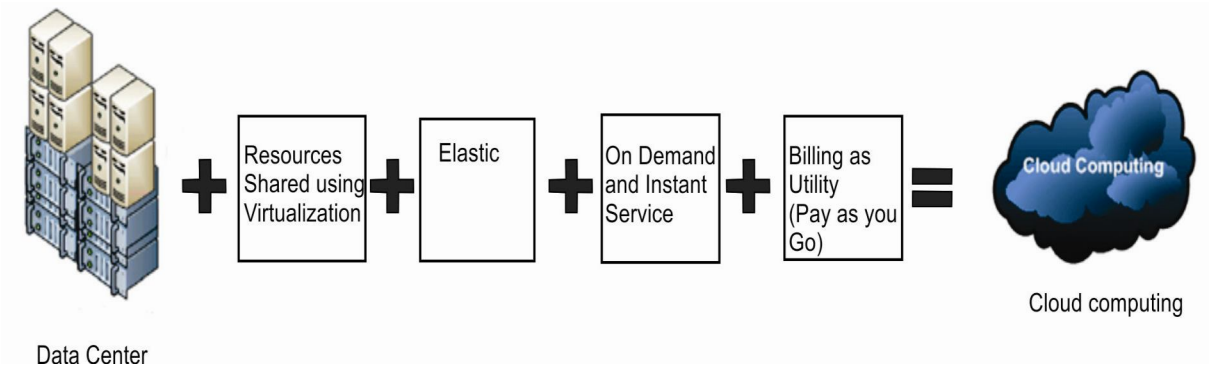

**Figure 2.** Schematic definition of cloud computing. Adapted from [6].

The major issues faced by Cloud Computing Security can be categorized as follows:

*A. Abuse and nefarious use of cloud computing:*

Some IaaS providers do not maintain enough security levels so that hackers and spammers can make use of this opportunity. A strict registration and identity checks would be needed, but still privacy laws a serious hindrance.

*B. Insecure application programming interfaces:*

Cloud providers supply some kind of software interfaces for the customers, weak and user friendly interfaces exposes security issues. The remedy would be strong authentication and access control with encrypted transmission.

*C. Malicious insiders:*

Higher level of access to an employee can leads to leakage of confidential data. The solution would be strict supply chain management and management practices which include controlling of privileged access. The best practices to handle this situation is using separation of privileges, least privilege, access control systems, alarm systems, administer logging, two factor authentication, codes of conduct, confidentiality agreements, background checks and visitor access.

*D. Shared technology vulnerabilities:*

Shared on-demand nature of the cloud computing needs virtualization and this technology is being used by the hypervisor to create virtual machines and operating systems. Any flaws in such hypervisor generate inappropriate access of the platform

*E. Data loss/leakage:*

Deletion or alteration of records without proper backups and loss of encoding key make the cloud difficult to restore. Unauthorized access into cloud can leads to data theft and losses.

*F. Account, service and traffic hijacking:*

Stolen credentials used for this kind attacks on the clouds which are usually taken by phishing, fraud or Denial of Services (DoS). The recommendations would be prohibition of sharing account credentials and two factor authentication techniques.

*G. Unknown risk profile:*

The reduction of hardware and software may not ensure the security in the cloud computing. An unknown risk profile is the cloud providers unwillingness to provide security logs, audit report and security practices.

*H. Information Security:*

Security related to the information exchanged between different hosts or between hosts and users. This issues pertaining to secure communication, authentication, and issues concerning single sign on and delegation. Secure communication issues include those security concerns that arise during the communication between two entities. These include confidentiality and integrity issues. Confidentiality indicates that all data sent by users should be accessible to only ”legitimate” receivers, and integrity indicates that all data received should only be sent/modified by ”legitimate” senders. Solution: public key encryption, X.509 certificates, and the Secure Sockets Layer (SSL) enables secure authentication and communication over computer networks.

## IV. PROPOSED APPROACH

*A. Motivation*

One of the existing solution deals with the protection of data with cryptographic key. But there is a chance of theft or getting the key by another person and behave like the owner. In order to avoid that, encrypt the key at the time when it is generated or periodically change the key. When the key generated is changed later, then it cause some difficulty for the owner. So at the same time encrypt the data also and this encryption can be done using various encryption algorithms like RSA, Blowfish etc. This can completely prevent the damage of data.

*B .Method of implementation*

Data should always be encrypted when stored (using separate symmetric encryption keys) and transmitted. If this is implemented appropriately, even if another tenant can access the data, all that will appear is gibberish. Shared symmetric keys for data encryption should be discouraged, yet tenants should be able to access their own encryption keys and change them when necessary. Cloud providers should not have ready access to tenant encryption keys. Data is not persistent in local system. So a Storage account is created with a cryptographic key. This storage account consists of container, Table, Queue. The container has a feature called blob, which is similar to files in Windows.

Blob is created using url:mystorage.blob.core.windows.net/my blobs/my blob. To access the storage once again, we have to use the URL and the cryptographic key. In order to prevent the loss of the key we have to frequently update and change the key. So a method is proposed such that the key and the data have to be encrypted before transmission. For this encryption we can use RSA algorithm or any other algorithms.

## V. CONCLUSION

Cloud computing is clearly one of today's most enticing technology areas due, at least in part, to its cost-efficiency and flexibility. The clouds have different architecture based on the services they provide. The data is stored on to centralized location called data centers having a large size of data storage. The data as well as processing is somewhere on servers. So, the clients have to trust the provider on the availability as well as data security. Before moving data into the public cloud, issues of security standards and compatibility must be addressed. A trusted monitor installed at the cloud server that can monitor or audit the operations of the cloud server. In minimizing potential security trust issues as well as adhering to governance issues facing Cloud computing, a prerequisite control measure is to ensure that a concrete Cloud computing Service Level Agreement (SLA) is put in place and maintained when dealing with outsourced cloud service providers and specialized cloud vendors. Cloud computing promises to change the economics of the data center, but before sensitive and regulated data move into the public cloud, issues of security standards and compatibility must be addressed including strong authentication, delegated authorization, key management for encrypted data, data loss protections, and regulatory reporting.

**REFERENCES**

[1] Armbrust, M., Fox, A., Griffith, R., Joseph, A. D., Katz, R., Konwinski, A., Lee, G., Patterson, D., Rabkin, A., Stoica, I., & Zaharia, M. (2010). A view of cloud computing. *Communications of the ACM*,*53*(4), 50-58.

[2] Buyya, R., Yeo, C. S., Venugopal, S., Broberg, J., & Brandic, I. (2009). Cloud computing and emerging IT platforms: Vision, hype, and reality for delivering computing as the 5th utility. *Future Generation computer systems*, *25*(6), 599-616.

[3] Marston, S., Li, Z., Bandyopadhyay, S., Zhang, J., & Ghalsasi,A.(2011).Cloud computing-The business perspective . *Decision Support Systems*, *51*(1),176-189.

[4] Buyya, R., Yeo, C. S., & Venugopal, S. (2008, September). Market-oriented cloud computing: Vision, hype, and reality for delivering it services as computing utilities. In *High Performance Computing and Communications, 2008. HPCC'08. 10th IEEE International Conference on* (pp. 5-13). Ieee.

[5] M. Armbrust, A. Fox, R. Griffith, A.D. Joseph, R.H. Katz, A. Konwinski, G. Lee, D.A. Patterson, A. Rabkin, I. Stoica, M. Zaharia, Above the Clouds: A Berkeley View of cloud computing, University of California at Berkeley, 2009.

[6] Khorshed, M. T., Ali, A. B. M., & Wasimi, S. A. (2012). A survey on gaps, threat remediation challenges and some thoughts for proactive attack detection in cloud computing. *Future Generation Computer Systems*, *28*(6), 833-851.

[7] Richard Chow, Philippe Golle, Markus Jakobsson, Ryusuke Masuoka, Jesus Molina, "*Controlling data in the cloud* ", http://portal.acm.org/citation.cfm?id=1655020 2009

[8] Satchit Dokras, Bret Hartman, Tim Mathers, "*The Role of Security in Trustworthy Cloud Computing*", 2009

[9] Cong Wang, Qian Wang, and Kui Ren, Wenjing Lou, "*Ensuring Data Storage Security in Cloud Computing*", 2009 , Page(s): 1 – 9

[10] Ramgovind S, Eloff MM, Smith E, "*The Management of Security in Cloud Computing*", 2010 , Page(s): 1 – 7

[11] Kreimir Popovi, eljko Hocenski, "*Cloud computing security issues and challenges*", May 24-28, MIPRO, 2010

Proceedings of the 33rd International Convention , Page(s):344-349

[12] Balachandra Reddy Kandukuri ,RamakrishnaPaturi, Dr. Atanu Rakshit,”*Cloud Security Issues* ”, pp.517-520, 2009 IEEE International Conference on Services Computing,
2009

[13]”C*loud Security*”, infoavail@sungard.com, December 2009

[14] John Harauz ,Lori M. Kaufman,Bruce Potter,” *Data Security in the World of Cloud Computing* ” IEEE Security and Privacy July 2009. pp. 61-64

[15] Fox, A., Griffith, R., Joseph, A., Katz, R., Konwinski, A., Lee, G., & Stoica, I. (2009). Above the clouds: A Berkeley view of cloud computing. *Dept. Electrical Eng. and Comput. Sciences, University of California, Berkeley, Rep. UCB/EECS*, *28*.

[6] Wired.com, The Future of cloud computing: A Long-TermForecast[cited 2009May 15];Available from: http://www.portfolio.com/views/columns/dual-perspectives/2009/03/09/A-Long-Term-Forecast/.